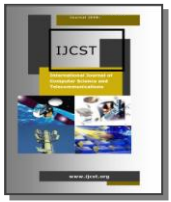

# A Machine Learning Model for Stock Market Prediction

Osman Hegazy [1], Omar S. Soliman [2] and Mustafa Abdul Salam [3]

[1,2] Faculty of Computers and Informatics, Cairo University, Egypt
[3] Higher Technological Institute (H.T.I), 10th of Ramadan City, Egypt
[2] dr_omar_soliman@yahoo.com, [3] mustafa.abdo@ymail.com

*Abstract*– Stock market prediction is the act of trying to determine the future value of a company stock or other financial instrument traded on a financial exchange. The successful prediction of a stock's future price will maximize investor's gains. This paper proposes a machine learning model to predict stock market price. The proposed algorithm integrates Particle swarm optimization (PSO) and least square support vector machine (LS-SVM). The PSO algorithm is employed to optimize LS-SVM to predict the daily stock prices. Proposed model is based on the study of stocks historical data and technical indicators. PSO algorithm selects best free parameters combination for LS-SVM to avoid over-fitting and local minima problems and improve prediction accuracy. The proposed model was applied and evaluated using thirteen benchmark financials datasets and compared with artificial neural network with Levenberg-Marquardt (LM) algorithm. The obtained results showed that the proposed model has better prediction accuracy and the potential of PSO algorithm in optimizing LS-SVM.

## I.　INTRODUCTION

STOCK price prediction has been at focus for years since it can yield significant profits. Predicting the stock market is not a simple task, mainly as a consequence of the close to random-walk behavior of a stock time series. Fundamental and technical analyses were the first two methods used to forecast stock prices. Artificial Neural networks (ANNs) is the most commonly used technique [1]. In most cases ANNs suffer from over-fitting problem due to the large number of parameters to fix, and the little prior user knowledge about the relevance of the inputs in the analyzed problem [2].

Also, Support vector machines (SVMs) had been developed as an alternative that avoids such limitations. Their practical successes can be attributed to solid theoretical foundations based on VC-theory [3]. SVM compute globally optimal solutions, unlike those obtained with ANNs, which tend to fall into local minima [4].

Least squares –support vector machines (LS-SVM) method was presented in [5], which was reformulated the traditional SVM algorithm. LS-SVM uses a regularized least squares function with equality constraints, leading to a linear system which meets the Karush-Kuhn-Tucker (KKT) conditions for obtaining an optimal solution. Although LS-SVM simplifies the SVM procedure, the regularization parameter and the kernel parameters play an important role in the regression system. Therefore, it is necessary to establish a methodology for properly selecting the LS-SVM free parameters, in such a way that the regression obtained by LS-SVM must be robust against noisy conditions, and it does not need priori user knowledge about the influence of the free parameters values in the problem studied [6].

The perceived advantages of evolutionary strategies as optimization methods motivated some researchers to consider such stochastic methods in the context of optimizing SVM. A survey and overview of evolutionary algorithms (EAs) found in [7]. Particle swarm optimization (PSO) is one of the most used EAs. PSO is a recently proposed algorithm by James Kennedy and Russell Eberhart in 1995, motivated by social behavior of organisms such as bird flocking and fish schooling [8]. The optimizer which is used in the particle swarm optimization algorithm, while making adjustment towards "local" and "global" best particles, is conceptually similar to the crossover operation used by genetic algorithms [9]. As well particle swarm optimization includes fitness function, which measures the closeness of the corresponding solution to the optimum. The main difference of particle swarm optimization concept from the evolutionary computing is that flying potential solutions through hyperspace are accelerating toward "better" solutions, while in evolutionary computation schemes operate directly on potential solutions which are represented as locations in hyperspace [10]. SVM was used in stock market forecasting in [11]. Financial time series forecasting using SVM optimized by PSO was presented in [12]. The Optimization of Share Price Prediction Model Based on Support Vector Machine is presented in [14]. Financial time series forecasting based on wavelet kernel support vector was presented in [15]. Computational Intelligence Approaches for Stock Price Forecasting was introduced in [16]. A hybrid approach by integrating wavelet-based feature extraction with MARS and SVR for stock index forecasting was presented in [17]. An interval type-2 Fuzzy Logic based system for modeling generation and summarization of arbitrage opportunities in stock markets was presented in [18]. Robust stock trading using fuzzy decision trees is presented in [19]. Ensemble ANNs-PSO-GA Approach for Day-ahead Stock E-exchange Prices Forecasting was presented in [20]. Index prediction with neuro-genetic hybrid network was proposed in [21]. A hybrid fuzzy intelligent agent-based system for stock price prediction was introduced in [22]. Improved Stock Market Prediction by Combining Support Vector Machine and Empirical Mode Decomposition was presented in [23]. Neural Network Ensemble Model Using PPR and LS-SVR for Stock Market Forecasting was proposed in [24]. Computational Intelligence Techniques for Risk Management in Decision Making was introduced in [25]. Stock market prediction algorithm using Hidden Markov Models was proposed presented in [26]. Neural Networks and Wavelet De-Noising for Stock Trading and Prediction was introduced in [27].

The aim of this paper is to develop a machine learning model that hybrids the PSO and LS-SVM model. The performance of LS-SVM is based on the selection of free parameters C (cost penalty), ϵ (insensitive-loss function) and γ (kernel parameter). PSO will be used to find the best parameter combination for LS-SVM.

The paper is organized as follows: Section 2 presents the Least square support vector machine algorithm; The Particle swarm optimization algorithm is introduced in section 3; The proposed model and its implementation stock prediction are discussed in section 4 Section 5 introduces excremental results and discussions. Finally, section 6 is devoted to conclusions of the proposed model.

## II.   LEAST SQUARE SUPPORT VECTOR MACHINE

Least squares support vector machines (LS-SVM) are least squares versions of support vector machines (SVM), which are a set of related supervised learning methods that analyze data and recognize patterns, and which are used for classification and regression analysis. In this version one finds the solution by solving a set of linear equations instead of a convex quadratic programming (QP) problem for classical SVMs. Least squares SVM classifiers, were proposed by Suykens and Vandewalle [28].

Let X is $n \times p$ input data matrix and $y$ is $n \times 1$ output vector. Given the $\{x_i, y_i\}_{i=1}^n$ training data set, where $xi \in R^p$ and $y_i \in R$, the LS-SVM goal is to construct the function $f(x) = y$, which represents the dependence of the output $y_i$ on the input $x_i$. This function is formulated as:

$$f(x) = W^T \varphi(x) + b \qquad (1)$$

Where $W$ and $\varphi(x)$: $R^p \to R^n$ are $n \times 1$ column vectors, and $b \in R$. LS-SVM algorithm [5] computes the function (1) from a similar minimization problem found in the SVM method [3]. However the main difference is that LS-SVM involves equality constraints instead of inequalities, and it is based on a least square cost function. Furthermore, the LS-SVM method solves a linear problem while conventional SVM solves a quadratic one. The optimization problem and the equality constraints of LS-SVM are defined as follows:

$$\min_{w,e,b} j(w,e,b) = \frac{1}{2} w^T w + C \frac{1}{2} e^T \qquad (2)$$

$$y_i = w^T \varphi(x_i) + b + e_i \qquad (3)$$

Where e is the $n \times 1$ error vector, 1 is a $n \times 1$ vector with all entries 1, and $C \in R^+$ is the tradeoff parameter between the solution size and training errors. From (2) a Lagranian is formed, and differentiating with respect to $w, b, e, a$ ($a$ is Largrangian multipliers), we obtain;

$$\begin{bmatrix} I & 0 & 0 & -Z^T \\ 0 & 0 & 0 & -1^T \\ 0 & 0 & CI & -I \\ Z & 1 & I & 0 \end{bmatrix} \begin{bmatrix} W \\ b \\ e \\ a \end{bmatrix} = \begin{bmatrix} 0 \\ 0 \\ 0 \\ y \end{bmatrix} \qquad (4)$$

Where $I$ represents the identity matrix and

$$Z = [\varphi(x1), \varphi(x2), ..., \varphi(x_n)]^T \qquad (5)$$

From rows one and three in (3) $w = Z^T a$ and $Ce = a$

Then, by defining the kernel matrix $K = ZZ^T$, and the parameter $\lambda = C^{-1}$, the conditions for optimality lead to the following overall solution

$$\begin{bmatrix} 0 & 1^T \\ 1 & K + \lambda I \end{bmatrix} \begin{bmatrix} b \\ a \end{bmatrix} = \begin{bmatrix} 0 \\ y \end{bmatrix} \qquad (6)$$

*Kernel function K types are as follows:*

- Linear kernel $K(x, x_i) = x_i^T x$     (7)

- Polynomial kernel of degree d:
$$K(x, x_i) = (1 + x_i^T x / c)^d \qquad (8)$$

- Radial basis function RBF kernel :

$$K(x, x_i) = \exp(-\|x - x_i\|^2 / \sigma^2) \qquad (9)$$

- MLP kernel :

$$K(x, x_i) = \tanh(k x_i^T x + \theta) \qquad (10)$$

In this work, MLP kernel is used.

## III.   PARTICLE SWARM OPTIMIZATION ALGORITHM

PSO is a relatively recent heuristic search method which is derived from the behavior of social groups like bird flocks or fish swarms. PSO moves from a set of points to another set of points in a single iteration with likely improvement using a combination of deterministic and probabilistic rules. The PSO has been popular in academia and industry, mainly because of its intuitiveness, ease of implementation, and the ability to effectively solve highly nonlinear, mixed integer optimization problems that are typical of complex engineering systems. Although the "survival of the fittest" principle is not used in PSO, it is usually considered as an evolutionary algorithm. Optimization is achieved by giving each individual in the search space a memory for its previous successes, information about successes of a social group and providing a way to incorporate this knowledge into the movement of the individual.

Therefore, each individual (called particle) is characterized by its position $\vec{x}_i$, its velocity $\vec{v}_i$, its personal best position $\vec{p}_i$ and its neighborhood best position $\vec{p}_g$.

The elements of the velocity vector for particle i are updated as:

$$\upsilon_{ij} \leftarrow \omega\upsilon_{ij} + c_1 q(x_{ij}^{pb} - x_{ij}) + c_2\gamma(x_j^{sb} - x_{ij}), j = 1,...,n \qquad (11)$$

Where w is the inertia weight, $x_i^{pb}$ is the best variable vector encountered so far by particle $i$, and $x^{sb}$ is the swarm best vector, i.e. the best variable vector found by any particle in the swarm, so far $c_1$ and $c_2$ are constants, and $q$ and $r$ are random numbers in the range (0, 1). Once the velocities have been updated, the variable vector of particle $i$ is modified according to:

$$x_{ij} \leftarrow x_{ij} + \upsilon_{ij}, j = 1,...,n. \qquad (12)$$

The cycle of evaluation followed by updates of velocities and positions (and possible update of $x_i^{pb}$ and $x^{sb}$) is then repeated until a satisfactory solution has been found. PSO algorithm is shown in Fig. 1.

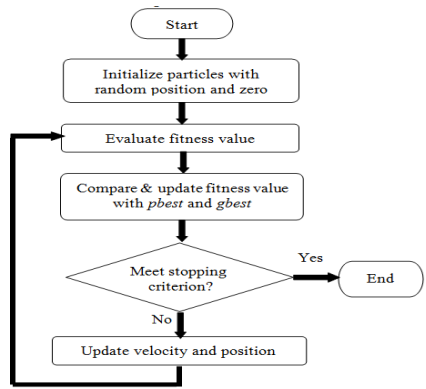

Fig. 1: PSO Algorithm

## IV. THE PROPOSED MODEL

The proposed model is based on the study of historical data, technical indicators and optimizing LS-SVM with PSO algorithm to be used in the prediction of daily stock prices. Levenberg-Marquardt (LM) algorithm is used as a benchmark for comparison with LS-SVM and LS-SVM-PSO models. The proposed model architecture contains six inputs vectors represent the historical data and derived technical indicators and one output represents next price.

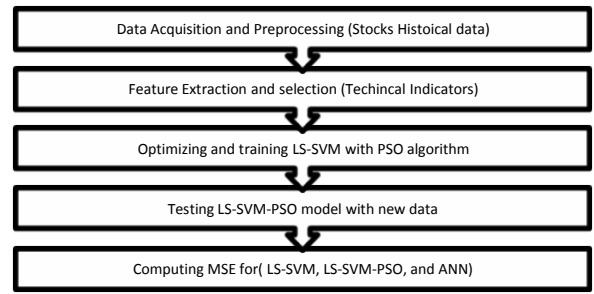

Fig. 2: The Proposed Model

The proposed algorithm was tested for many companies which cover all stock sectors in S&P 500 stock market. These sectors are Information Technology (Adobe, Hp, and Oracle); Financials (American Express and Bank of New York); Health Care (Life Technologies, and Hospera); Energy (Exxon-Mobile and Duck energy); Communications (AT&T); Materials (FMC Corporation); Industrials (Honey Well).

*Five technical indicators are calculated from the raw datasets:*

- *Relative Strength Index (RSI)*: A technical momentum indicator that compares the magnitude of recent gains to recent losses in an attempt to determine overbought and oversold conditions of an asset. The formula for computing the Relative Strength Index is as follows.

   RSI = 100- [100 / (1+RS)]                    (13)

   Where RS = Avg. of x days' up closes / Average of x days' down closes.

- *Money Flow Index (MFI):* This one measures the strength of money in and out of a security. The formula for MFI is as follows:

   Money Flow (MF) = Typical Price * Volume.        (14)

   Money Ratio (MR) = (Positive MF / Negative MF).    (15)

   MFI = 100 − (100/ (1+MR)).                    (16)

- *Exponential Moving Average (EMA):* This indicator returns the exponential moving average of a field over a given period of time. EMA formula is as follows.

   EMA = [α *T Close] + [1-α* Y EMA].            (17)

   Where T is Today's close and Y is Yesterday's close

- *Stochastic Oscillator (SO):* The stochastic oscillator defined as a measure of the difference between the current closing price of a security and its lowest low price, relative to its highest high price for a given period of time. The formula for this computation is as follows:

   %K = [(Close price − Lowest price) / (Highest Price − Lowest Price)] * 100            (18)

- *Moving Average Convergence/Divergence (MACD):* This function calculates difference between a short and a long term moving average for a field. The formulas for calculating MACD and its signal as follows.

MACD = [0.075*EMA of Closing prices] −
$\qquad$ [0.15*EMA of closing prices] $\qquad$ (19)

Signal Line = 0.2*EMA of MACD $\qquad$ (20)

## V.    RESULTS AND DISCUSSION

LS-SVM-PSO, LS-SVM and ANN algorithms were trained and tested with datasets form Jan 2009 to Jan 2012. All datasets are available in [13]. All datasets are divided into training part (70%) and testing    part (30%).

Fig. 3 to Fig. 14 outline the application of Proposed LS-SVM-PSO model compared with LS-SVM and ANN-BP algorithms at different data set with different sectors of the market. In Fig. 3, Fig. 4, and Fig. 5, which present results of three companies in information technology sector (Adobe, Oracle and HP), results show that LS-SVM optimized with PSO is the best one with lowest error value followed by LS-SVM algorithm. Fig. 6 and Fig. 7 represent results of financial sector (American Express, and Bank of New york), we can remark that the predicted curve using the proposed LS-SVM-PSO algorithm is most close to the real curve which achieves best accuracy, followed by LS-SVM, while    ANN-BP is the worst one.

Fig. 8 represents results of using PSO-LS-SVM model in Honeywell company which represent industrials stock sector, proposed model still achieves best performance.

Fig. 9 and Fig. 10 outline the application of the proposed algorithm to hospera and life technologies companies in health stock sector. From figures one can remark the enhancement in the error rate achieved by the proposed model.

Fig. 11 and Fig. 12 outline the results of testing proposed model on Exxon-mobile and duke energy companies which represent energy stock sector. PSO-LS-SVM also the best especially in fluctuation cases.

Fig. 13 represents results for FMC Corporation in materials stock sector. The achievements of proposed model is very promising compared with SVM and ANN

Fig. 14 outlines results for AT&T from communication stock sector. We can notice from figure the role of proposed model in reducing the error rate and overcoming local minima problems which found in ANN results.

Table 1 outlines Mean Square Error (MSE) performance function for proposed algorithm. It can be remarked that the LS-SVM optimized with PSO always gives an advance over LS-SVM and ANN trained with LM algorithms in all performance functions and in all trends and sectors. Proposed model performs better than other algorithms especially in cases with fluctuations in the time series function.

Fig. 15 outlines comparison between PSO-LS-SVM, LS-SVM and ANN algorithms according to MSE function.

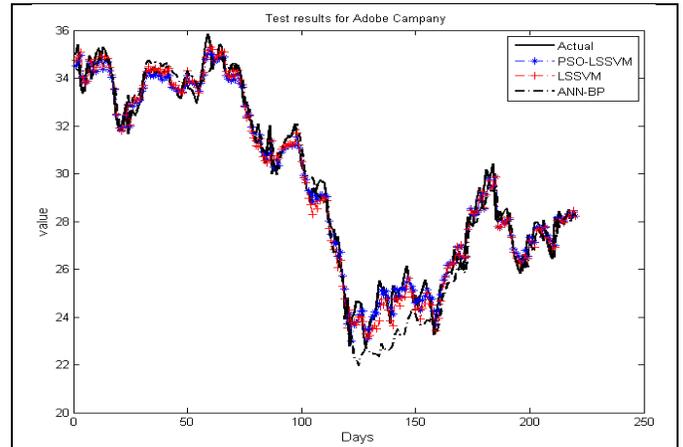

Fig. 3: Results for Adobe Company

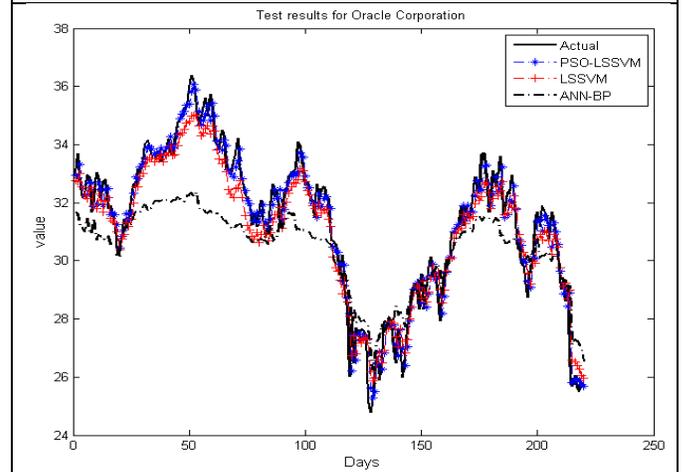

Fig. 4: Results for Oracle Company

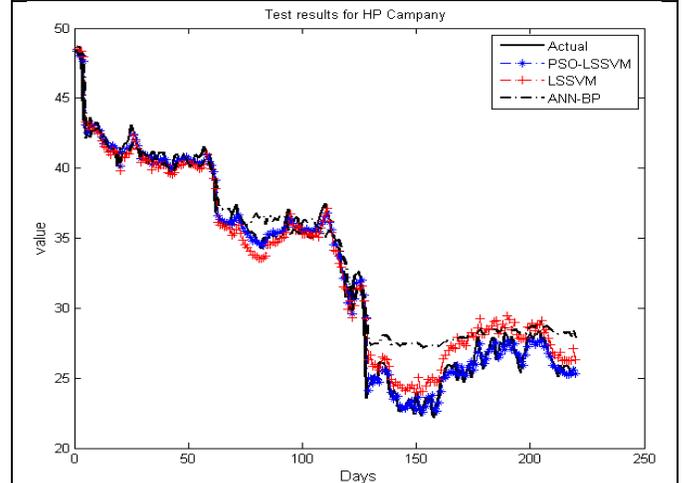

Fig. 5: Results for HP Company

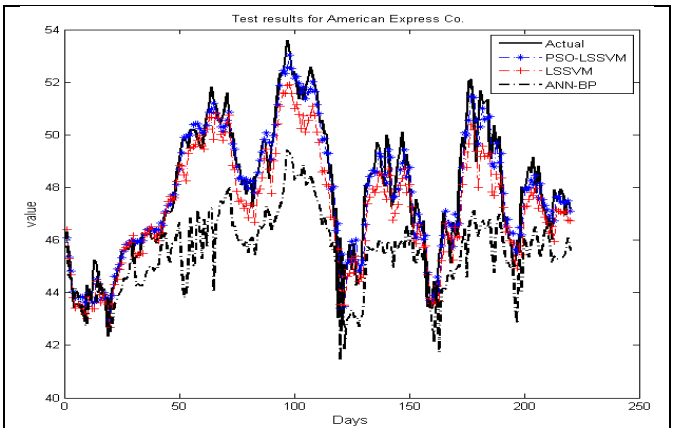

Fig. 6: Results for American Express Co

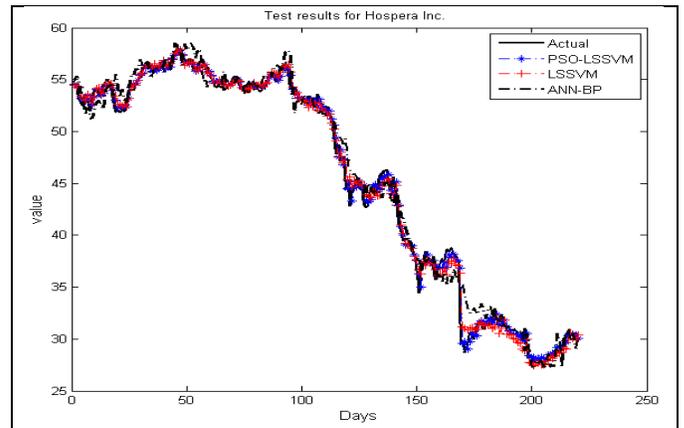

Fig. 9: Results for Hospera Company

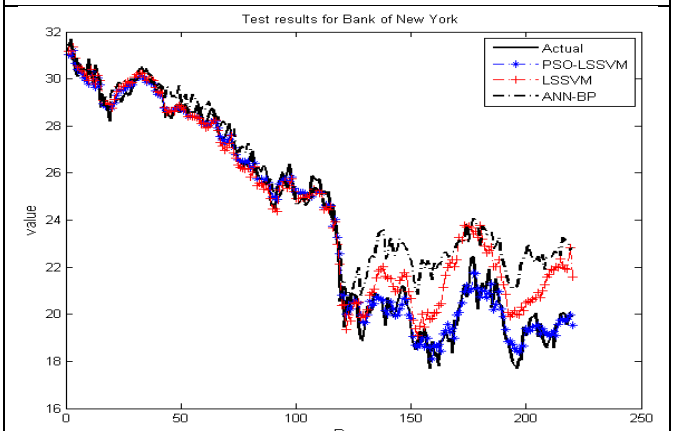

Fig. 7: Results for Bank of New York

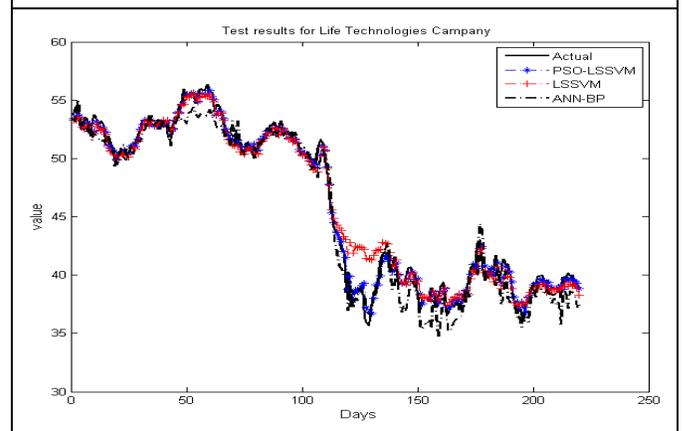

Fig. 10: Results for Life Technologies company

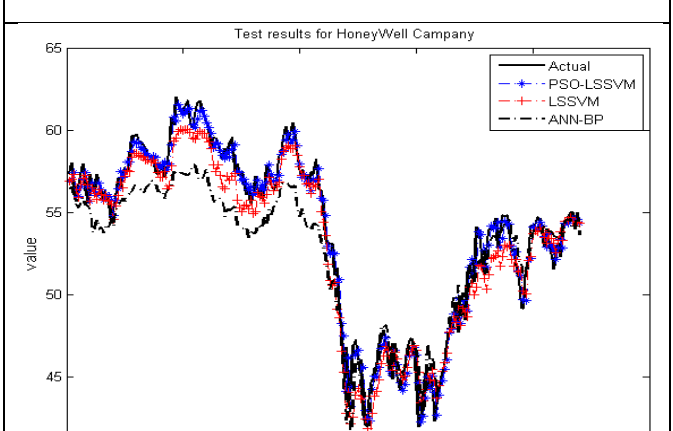

Fig. 8: Results for Honeywell company

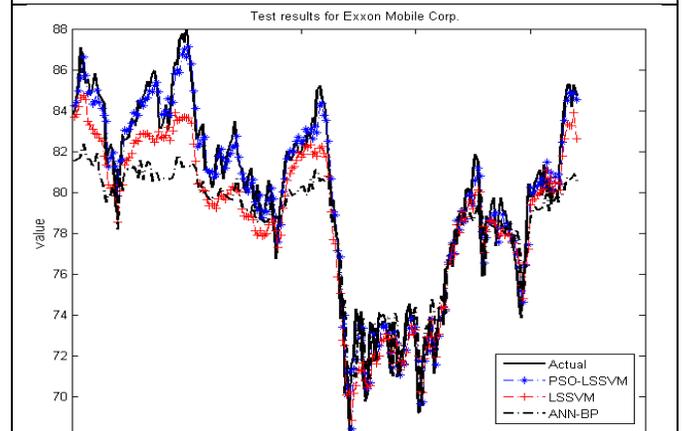

Fig. 11: Results for Exxon Mobile company

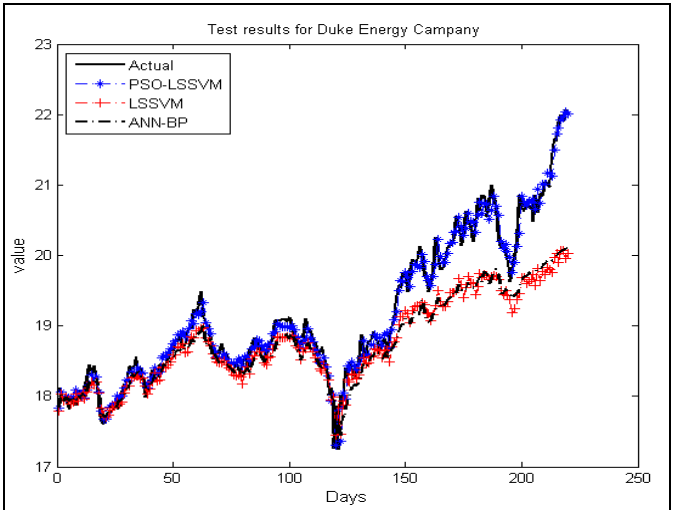

Fig. 12: Results for Duke Energy company

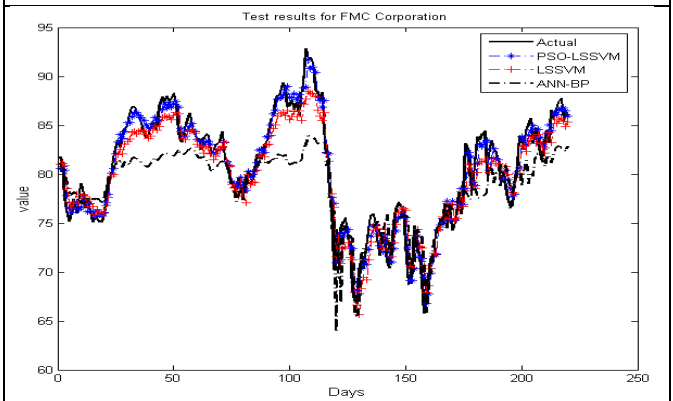

Fig. 13: Results for FMC corporation

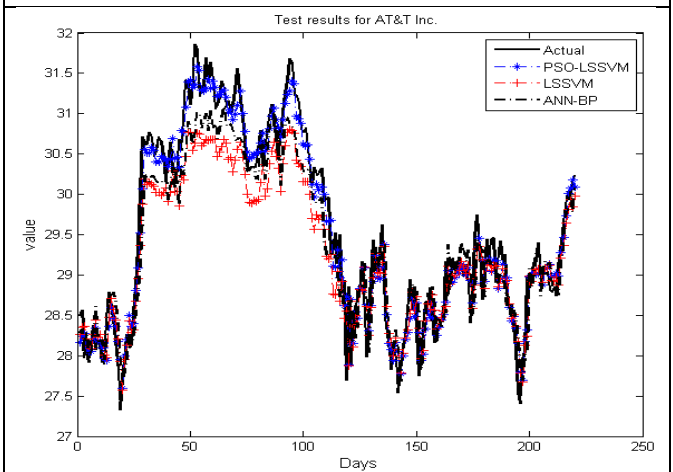

Fig. 14: Results for AT &T Company

Table 1: Mean Square Error (MSE) for proposed algorithm

| Algorithm / Company | PSO-LS-SVM | LS-SVM | NN-BP |
|---|---|---|---|
| Adobe | 0.5317 | 0.5703 | 0.8982 |
| Oracle | 0.6314 | 0.8829 | 0.9124 |
| HP | 0.7725 | 1.2537 | 1.9812 |
| American Express | 0.7905 | 1.0663 | 2.8436 |
| Bank of New york | 0.4839 | 1.2769 | 1.9438 |
| Coca-Cola | 0.6823 | 0.9762 | 1.7975 |
| HoneyWell | 0.9574 | 1.3371 | 2.1853 |
| Hospera | 0.8694 | 0.9320 | 1.4640 |
| Life Tech. | 0.7713 | 1.3221 | 1.3492 |
| Exxon-Mobile | 1.1000 | 1.6935 | 2.4891 |
| AT & T | 0.2911 | 0.4673 | 0.4055 |
| FMC Corp. | 1.5881 | 2.1034 | 3.5049 |
| Duke Energy | 0.1735 | 0.6097 | 0.6010 |

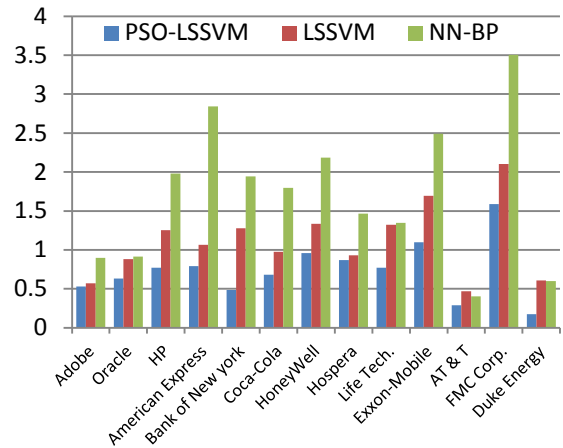

Fig. 15: MSE of proposed algorithm

## VI.    CONCLUSIONS

This paper, proposed a machine learning model that integrates particle swarm optimization (PSO) algorithm and LS-SVM for stock price prediction using financial technical indicators. These indicators include relative strength index, money flow index, exponential moving average, stochastic oscillator and moving average convergence/divergence The PSO is employed iteratively as global optimization algorithm to optimize LS-SVM for stock price prediction. Also, PSO algorithm used in selection of LS-SVM free parameters C (cost penalty), $\epsilon$ (insensitive-loss function) and $\gamma$ (kernel parameter). The proposed LS-SVM-PSO model convergence to the global minimum. Also, it is capable to overcome the over-fitting problem which found in ANN, especially in case of fluctuations in stock sector. PSO-LS-SVM algorithm parameters can be tuned easily. The performance of the proposed model is better than LS-SVM and compared algorithms. LS-SVM-PSO achieves the lowest error value followed by single LS-SVM, while ANN-BP algorithm is the worst one.